\documentstyle[11pt]{article}
\voffset -0.6in \hoffset -0.6in {\setlength{\oddsidemargin}{0.8in}
\setlength{\evensidemargin}{0.8in}

\begin{document}
\baselineskip 13pt

\title{
On the Variable-charged Black Holes Embedded into de Sitter Space:
Hawking's Radiation}

\author{Ng. Ibohal \\
Department of Mathematics, Manipur University,\\
Imphal 795003, Manipur, INDIA.\\
E-mail:  (i)
ngibohal@rediffmail.com \\
(ii) ngibohal@iucaa.ernet.in }
\date{May 4, 2004}

\maketitle

\begin{abstract}
In this paper we discuss the Hawking's evaporation of the masses
of variable-charged Reissner-Nordstrom and Kerr-Newman, black
holes embedded into the de Sitter cosmological universe by
considering the charge to be function of radial coordinate. It has
been shown that every electrical radiation of variable-charged
rotating or non-rotating cosmological black holes will produce a
change in the mass of the body without effecting the Maxwell
scalar and the cosmological constant. It is also shown that during
the Hawking's radiation process, after the complete evaporation of
masses of both variable-charged Reissner-Nordstrom-de Sitter and
Kerr-Newman-de Sitter black holes, the electrical radiation will
continue creating negative mass naked singularities
embedded into de Sitter cosmological spaces 
\end{abstract}

\section{Introduction}

The Hawking's radiation [1] suggests that black holes which are
formed by collapse, are not completely black, but emit radiation
with a thermal spectrum. Hawking [2] stated that `Because the
radiation carries away energy, the black holes must presumably
lose mass and eventually disappear'. In an introductory survey
Hawking and Israel [3] have discussed the black hole radiation in
three possible ways with creative remarks --`So far there is no
good theoretical frame work with which to treat the final stages
of a black hole but there seem to be three possibilities: (i) The
black hole might disappear completely, leaving just the thermal
radiation that it emitted during its evaporation. (ii) It might
leave behind a non-radiating black hole of about the Planck mass.
(iii) The emission of energy might continue indefinitely creating
a negative mass naked singularity'. In [4] it has shown that
Hawking's radiation could be expressed in classical spacetime
metrics, by considering the charge $e$ of the electromagnetic
field to be function of the radial coordinate $r$ of
Reissner-Nordstrom as well as Kerr-Newman black holes.  Such a
variable charge $e$ with respect to the coordinate $r$ in
Einstein's equations is referred to as {\it an electrical
radiation} of the black hole. Every electrical radiation e(r) of
the non-rotating as well as rotating black holes leads to a
reduction in its mass by some quantity. If one considers such
electrical radiation taking place continuously for a long time,
then a continuous reduction of the mass will take place in the
black hole body whether rotating or non-rotating, and the original
mass of the black hole will evaporate completely. At that stage
the complete evaporation will lead the gravity of the object
depending only on the electromagnetic field, and not on the mass.
We refer to such an object with zero mass as an `instantaneous'
naked singularity - a naked singularity that exists for an instant
and then continues its electrical radiation to create negative
mass. So this naked singularity is different from the one
mentioned in Steinmular {\it et al.} [5], Tipler {\it et al.} [6]
in the sense that an `instantaneous' naked singularity, discussed
in [5,6] exists only for an instant and then disappears.

We note that the time taken between two consecutive radiations is
supposed to be so short that one may not physically realize how
quickly radiations take place. Thus, it seems natural to expect
the existence of an `instantaneous' naked singularity with zero
mass only for an instant before continuing its next radiation to
create a negative mass naked singularity. This suggests that it
may also be possible in the common theory of black holes that, as
a black hole is invisible in nature, one may not know whether, in
the universe, a particular black hole has mass or not, but
electrical radiation may be detected on the black hole surface.
Immediately after the complete evaporation of the mass, if one
continues to radiate the remaining remnant, there will be a
formation of a new mass. If one repeats the electrical radiation
further, the new mass will increase gradually and then the
spacetime geometry will represent the `negative mass naked
singularity'. The classical spacetime metrics, for both stationary
rotating and non-rotating, which represent the negative mass naked
singularities, have been presented in [4]. Here the variable
charge $e(r)$ with respect to the coordinate $r$ is followed from
Boulware's suggestion [7] that the Hawking's radiation may be
expressed in terms of the stress-energy tensor associated with
field whose quanta are being radiated. In order to study Hawking's
radiation in classical spacetime metrics, the Boulware's
suggestion leads us to consider the stress-energy tensors of
electromagnetic field of different forms or functions from those
of Reissner-Nordstrom, as well as Kerr-Newman, black holes as
these two black holes do not seem to have any direct Hawking's
radiation effects. Thus, we find that (i) the changes in the mass
of black holes, (ii) the formation of `instantaneous' naked
singularities  with zero mass and (iii) the creation of `negative
mass naked singularities' in Reissner-Nordstrom as well as
Kerr-Newman black holes [4] may presumably  be the correct
formulations in classical spacetime metrics of the three
possibilities of black hole evaporation suggested by Hawking and
Israel [3].

The aim of this paper is to study the relativistic aspect of
Hawking's radiation in Reissner-Nordstrom-de Sitter  as well as
Kerr-Newman-de Sitter black holes by considering the variable
charge with respect to the coordinate $r$. The results are
summarized in the form of theorems as follows:
\newtheorem{theorem}{Theorem}
\begin{theorem}
Every electrical radiation of variable-charged
Reissner-\\Nordstrom-de Sitter and Kerr-Newman-de Sitter black
holes will produce a change in the mass of the bodies without
affecting the Maxwell scalar and the cosmological constant.
\end{theorem}
\begin{theorem}
The non-rotating and rotating charged de Sitter metrics describe
instantaneous de Sitter naked singularity during the Hawking's
evaporation process of electrical radiation of
Reissner-Nordstrom-de Sitter and Kerr-Newman-de Sitter black
holes.
\end{theorem}
\begin{theorem}
During the radiation process, after the complete evaporation of
masses of both variable-charged Reissner-Nordstrom-de Sitter and
Kerr-Newman-de Sitter black holes, the electrical radiation will
continue indefinitely creating negative mass naked singularities
in de Sitter spaces.
\end{theorem}
\begin{theorem}
If an electrically radiating black hole, rotating or non-rotating,
is embedded into  de Sitter spaces, it will continue to embed into
the same space forever.
\end{theorem}

It is found that the theorems 1, 2 and 3 are in favour of the
first, second and third possibilities of the suggestions made by
Hawking and Israel [3]. But theorem 3 provides a violation of
Penrose's cosmic-censorship hypothesis that `no naked singularity
can ever be created' [8]. Classical spacetime metrics describing
theorem 3 have been derived below. Theorem 4 suggests that, once
an electrically radiating black hole is embedded into the de
Sitter cosmological universe, in principle it will continue to
embed forever during its radiation process. It happens because it
does not seem to have any possible mathematical method to remove
the cosmological constant from the Einstein's field equations of
radiating universe, unless some external forces apply to it.

Here, we shall use the phrase `change in the mass' rather then
`loss of mass' as there is a possibility of creating new mass
after the exhaustion of the original mass, if one repeats the same
process of electrical radiation. This can be seen latter in this
paper. Hawking's radiation is being incorporated, in the classical
general relativity describing the change in mass appearing in the
classical space-time metrics, without quantum mechanical aspect as
done by Hawking [1] or the path integral method used by Hurtle and
Hawking [9] or thermodynamic viewpoint [10]. In section 2 we
present classical spacetime metrics affected by the  change in the
mass of the `variable-charged' black holes embedded into de Sitter
spaces after electrical radiation. In Section 3 the properties of
the metrics formed after the electrical radiation are being
discussed with regards to the Kerr-Schild form and Chandrasekhar's
relation [8]. The NP (Newman and Penrose [11]) version of original
rotating de Sitter, Reissner-Nordstrom-de Sitter and
Kerr-Newman-de Sitter solutions [12] are presented in the
appendices. The NP quantities are calculated by using the
differential form structure in NP formalism developed by McIntosh
and Hickman [13] in (--2) signature.

\section{Changing masses of variable-charged black holes
embedded in de Sitter space}

\setcounter{equation}{0}
\renewcommand{\theequation}{2.\arabic{equation}}

In this section we shall consider the variable-charged black holes
embedded into de Sitter space with the charge $e(r)$. By solving
Einstein-Maxwell field equations with the variable-charge $e(r)$
of Reissner-Nordstrom as well as Kerr-Newman, black holes embedded
in de Sitter spaces, we develop the relativistic aspect of Hawking
radiation in classical spacetime metrics. It is to mention that in
the formulation of the relativistic aspect of Hawking's radiation,
we do not impose any condition on the field equations except
considering the charge $e$ to be function of polar coordinate $r$
and the decomposition of the Ricci scalar
$\Lambda\equiv(1/24)\,R_{ab}\,g^{ab}$ into two parts, without loss
of generality, as follows
\begin{equation}
\Lambda=\Lambda^{(\rm C)}+\Lambda^{(\rm E)},
\end{equation}
where $\Lambda^{(\rm C)}$ is the {\it non-zero} cosmological Ricci
scalar, and $\Lambda^{(\rm E)}$ is the {\it zero} Ricci scalar of
electromagnetic field for the rotating as well as non-rotating
black holes, which can be seen in the equation (C9) of appendix
cited below. This decomposition of Ricci scalar $\Lambda$ is
possible because the cosmological object and the electromagnetic
field are two different matter fields of different physical
nature, though they are supposed to exist on the same spacetime
coordinates here. For our purpose of the paper, this type of
decomposition of Ricci scalars $\Lambda$ will serve well in the
study of Hawking's radiation of black holes embedded into the de
Sitter cosmological space.

\subsection{\sl Variable-charged Reissner-Nordstrom-de
Sitter solution}

The line element of the variable-charged Reissner-Nordstrom-de
Sitter solution with the assumption that the charge $e$ of the
body is a function of coordinate $r$, is given by

\begin{equation}
ds^2=\Big\{1-{2M\over r}+{e^2(r)\over
r^2}-{\Lambda^*r^2\over3}\Big\}\,du^2+2du\,dr-r^2(d\theta^2
+{\rm sin}^2\theta\,d\phi^2).
\end{equation}
When we set $e=$ constant initially, the metric will recover the
charged Reissner-Nordstrom-de Sitter solution given in the
appendix B. Here $\Lambda^*$ is the cosmological constant of the
de Sitter universe. The complex null tetrad vectors for this
metric are chosen as
\begin{eqnarray*}
\ell_a=\delta^1_a,
\end{eqnarray*}
\begin{equation}
n_a={1\over 2}\Big\{1-{2m\over
r}+{e^2\over r^2}-{\Lambda^*r^2\over
3}\Big\}\delta^1_a+\delta^2_a,
\end{equation}
\begin{eqnarray*}
m_a=-{r\over\surd 2}\,\{\delta^3_a+i\,{\rm
sin}\,\theta\,\delta^4_a\},
\end{eqnarray*}
\begin{eqnarray*}
\overline m_a=-{r\over\surd 2}\,\{\delta^3_a-i\,{\rm sin}\,\theta\,\delta^4_a\}.
\end{eqnarray*}
where $\ell_a$,\, $n_a$ are real null vectors and $m_a$ is complex
null vector. Using these null tetrad vectors we calculate the spin
coefficients, Ricci scalars and Weyl scalars as follows:
\begin{eqnarray*}
\kappa=\epsilon=\sigma=\nu=\lambda=\pi=\tau=0,\;\;\ \rho=-{1\over r},\;\
\beta=-\alpha={1\over {2\surd 2r}}\,{\rm cot}\theta,
\end{eqnarray*}
\begin{eqnarray*}
\mu=-{1\over 2\,r}\Big\{1-{2M\over r}-{\Lambda^*r^2\over
3}+{e^2(r)\over r^2}\Big\},\;\;\
\end{eqnarray*}
\begin{eqnarray*}
\gamma={1\over 2\,r^2}\,\Big\{M+e(r)\,e'(r)-{e^2(r)\over r}
-{\Lambda^*r^2\over
3}\Big\},\;\
\end{eqnarray*}
\begin{equation}
\phi_{11}={1\over 4\,r^2}\,\Big\{e'^2(r)+e(r)\,e''(r)\Big\}+
{1\over 2\,r^4}\Big\{e^2(r)-2\,r\,e(r)e'(r)\Big\},\;\
\end{equation}
\begin{equation}
\Lambda = {\Lambda^*\over 6}-{1\over
12\,r^2}\,\{e'^2(r)+e(r)\,e''(r)\},
\end{equation}
\begin{equation}
\psi_2={1\over 6\,r^2}\,\Big\{e'^2(r)+e(r)\,e''(r)\Big\}-{1\over
r^3}\,\Big\{M
+ e(r)e'(r)- {e^2(r) \over r}\Big\},
\end{equation}
where a prime denotes the derivative with respect to $r$.
The Weyl curvature scalar $\psi_2$, Ricci scalars $\phi_{11}$ and
$\Lambda$  for the metric (2.2) are defined by
\begin{eqnarray*}
\psi_2 \equiv -C_{abcd}\,\ell^a\,m^b\,\overline{m}^c\,n^d,
\end{eqnarray*}
\begin{equation}
\phi_{11}\equiv-{1\over4}\,R_{ab}(\ell^a\,n^b
+m^a\,\overline{m}^b),  \:\:\:\\\
\Lambda\equiv{1\over24}\,R_{ab}\,g^{ab}.
\end{equation}
We have seen from above that there is no $\Lambda^*$ term in
$\phi_{11}$ and $\psi_2$. So the Ricci scalar $\phi_{11}$ is
purely for electromagnetic field. Hence, using the decomposition
(2.1) of $\Lambda$ in equation (2.5) we obtain
\begin{equation}
\Lambda^{(\rm C)}= {\Lambda^*\over 6},
\end{equation}
and
\begin{equation}
\Lambda^{(\rm E)} = -{1\over 12\,r^2}\,\{e'^2(r)+e(r)\,e''(r)\},
\end{equation}
For an electromagnetic field we must have the Ricci scalar
$\Lambda^{(\rm E)}=0$ leading to the solution
\begin{equation}
e^2(r) = 2\,rm_1 + C
\end{equation}
where  $m_1$ and $C$ are real constants. Then the Ricci scalar
becomes
\begin{equation}
\phi_{11} ={C\over 2\,r^4}.
\end{equation}
Thus, the Maxwell scalar $\phi_1={1\over 2}
F_{ab}(\ell^a\,n^b+\overline{m}^a\,m^b)$ takes the form, by identifying the
real constant $C \equiv e^2$,
\begin{equation}
\phi_1={1\over\surd 2}\,e\,r^{-2}.
\end{equation}
This shows that the Maxwell scalar $\phi_1$ does not change its form
by considering the charge $e$ to be a function of $r$ in
Einstein-Maxwell field equations. Here, by using equation (2.10)
in (2.4) and (2.6), we have the changed NP quantities
\begin{equation}
\mu=-{1\over 2\,r}\Big\{1-{2\over r}(M - m_1)+{e^2\over
r^2}-{\Lambda^*r^2\over 3}\Big\},\;\;\
\end{equation}
\begin{eqnarray*}
\gamma={1\over 2\,r^2}\,\Big\{(M - m_1)-{e^2\over r}-{\Lambda^*r^2\over
3}\Big\},\;\
\end{eqnarray*}
\begin{equation}
\psi_2=-{1\over r^3}\,\Big\{(M - m_1)-{e^2\over r}\Big\},\;\;\
\phi_1={1\over\surd 2}\,e\,r^{-2},
\end{equation}
and the metric (2.2) becomes
\begin{equation}
ds^2=\Big\{1-{2\over r}(M - m_1)+{e^2\over
r^2}-{\Lambda^*r^2\over
3}\Big\}\,du^2+2du\,dr-r^2(d\theta^2
+{\rm sin}^2\theta\,d\phi^2).
\end{equation}
We observe from the above that the mass $M$ of non-rotating black
hole (2.2) is lost a quantity $m_1$ at the end of the first
electrical radiation. This loss of mass is agreeing with Hawking's
discovery that the radiating objects must lose its mass [2]. On
this losing mass, Wald [14] has pointed that a black hole will
lose its mass at the rate as the energy is radiated. If one
considers the same process for second time taking $e$ in (2.15) to
be function of $r$ with the mass $M - m_1$ in Einstein-Maxwell
field equations, then the mass will again be decreased by another
constant $m_2$ (say); that is, after the second radiation the
total mass might become $M - (m_1 + m_2)$. This is due to the
fact, that the Maxwell scalar $\phi_1$ does not change its form
after considering the charge $e$ to be function of $r$ for the
second time as $\Lambda^{(\rm E)}$ calculated from the
Einstein-Maxwell field equations has to vanish for electromagnetic
fields with $e(r)$. Hence, if one repeats the same process $n$
times, everytime considering the decomposition of $\Lambda$ and
the charge $e$ to be function of $r$, then one can expect the
solution to change gradually and the total mass becomes $M - (m_1
+ m_2 + m_3 + . . . + m_n)$ and therefore the metric (2.15) takes
the form:
\begin{equation}
ds^2=\Big\{1-{2\over r}\,{\cal M}+{e^2\over r^2}-{\Lambda^*r^2\over 3}
\Big\}\,du^2+2du\,dr-r^2(d\theta^2+{\rm sin}^2\theta\,d\phi^2),
\end{equation}
where the mass of the black hole after the radiation for $n$ times
will be
\begin{equation}
{\cal M} = M - (m_1 + m_2 + m_3 + . . . + m_n).
\end{equation}
So, the changed NP quantities are
\begin{equation}
\mu=-{1\over 2\,r}\Big\{1-{2\over r}{\cal M}+{e^2\over
r^2}-{\Lambda^*r^2\over 3}\Big\},
\end{equation}
\begin{equation}
\gamma={1\over 2\,r^2}\,\Big\{{\cal M}-{e^2\over r}-{\Lambda^*r^2\over
3}\Big\},
\end{equation}
\begin{equation}
\psi_2=-{1\over r^3}\,\Big\{{\cal M}-{e^2\over r}\Big\}.
\end{equation}

This means that for every electrical radiation, the original mass
$M$ of the non-rotating black hole (2.2) will lose some quantity.
Thus, it seems reasonable to expect that, taking Hawking's
radiation of black holes into account, such continuous lose of
mass will lead to evaporate the original mass $M$. In the case the
black hole has evaporated down to the Planck mass, the mass $M$
may not exactly equal to the continuously lost quantities $m_1 +
m_2 + m_3 + . . . + m_n$. That is, according to the second
possibility of Hawking and Israel [3], a small quantity of mass
may be left, say, Planck mass of about $10^{-5}$ g with continuous
electrical radiation. Otherwise, when $M$ = $m_1 + m_2 + m_3 + . .
. + m_n$ for a complete evaporation of the mass, ${\cal M}$ will
be zero, rather than leaving behind a Planck-size mass black hole
remnant. At this stage the non-rotating black hole will have the
electric charge $e$ and the cosmological constant $\Lambda^*$, but
no mass; so that the line element will take the following form:
\begin{equation}
ds^2=\Big(1+{e^2\over r^2}-{\Lambda^*r^2\over 3}\Big)\,du^2
+2du\,dr-r^2(d\theta^2+{\rm
sin}^2\theta\,d\phi^2).
\end{equation}
That is, the black hole embedded into de Sitter cosmological space
might radiate away  all its mass completely, just leaving the
electrical radiation and the unchanged cosmological constant
$\Lambda^*$. At this stage the gravity of the surface will depend
only on electric charge, {\it i.e.} $\psi_2=e^2/r^4$, and not on
the mass of black holes. The metric (2.21) describes a
non-rotating charged de Sitter cosmological solution. Thus, we
consider this non-rotating charged de Sitter solution to be the
left out remnant of the Hawking's evaporation due to the
electrical radiation of Reissner-Nordstrom black hole embedded
into the non-rotating de Sitter space. The metric (2.21) shows
that the presence of the cosmological constant $\Lambda^*$ during
Hawking's radiation process could not prevent the formation of an
`instantaneous' naked singularity. The formation of
`instantaneous' naked singularity - {\it a naked singularity that
exists for an instant and then continues its electrical radiation
to create negative mass}, in standard Reissner-Nordstrom and
Kerr-Newman, black holes is unavoidable during the Hawking's
evaporation process, as shown in [4]. That is, if we set the
cosmological constant $\Lambda^*=0$, the metric (2.21) will
certainly represent an `instantaneous' naked singularity with zero
mass, However, the Maxwell scalar $\phi$ is still unaffected.
Thus, from (2.21) with $\Lambda^*\neq 0$ it seems natural to refer
to the non-rotating charged cosmological metric as an
`instantaneous' naked singularity in de Sitter space - {\it a
singularity that exists for an instant and then continues its
electrical radiation to create its negative mass}, during the
Hawking's evaporation process of electrical radiation of
Reissner-Nordstrom-de Sitter black hole. This completes the proof
of the first part of theorem 2 cited in the introduction.

It is suggested that the time taken between two consecutive
radiations is supposed to be so short that we may not physically
realize how quickly radiations take place.  Thus it seems natural
to expect the existence of `instantaneous' naked singularity with
zero mass in de Sitter cosmological space only for an instant
before continuing its next radiation to create new mass.
Immediately,  after the exhaustion of the Reissner-Nordstrom mass,
if the remaining solution (2.21) continues to radiate electrically
with $e(r)$, there will be a formation of new mass $m^*_1$ (say).
If this electrical radiation process continues forever, the new
mass will increase gradually as
\begin{equation}
{\cal M}^*=m^*_1 + m^*_2 + m^*_3 + m^*_4 + . . .
\end{equation}
and then the metric with the new mass becomes
\begin{equation}
ds^2=\Big(1+{2\over r}\,{\cal M}^* + {e^2\over
r^2}-{\Lambda^*r^2\over 3}\Big)\,du^2+2du\,dr-r^2(d\theta^2 +{\rm
sin}^2\theta\,d\phi^2).
\end{equation}
However, it appears that this new mass ${\cal M}^*$ will never
decrease, rather it might increase gradually as the radiation
continues forever. Then the Weyl scalar for the metric (2.23)
becomes
\begin{equation}
\psi_2={1\over r^3}\,\Big\{{\cal M}^*+{e^2\over r}\Big\},\\\\
\end{equation}
which is different from the one given in (2.20) by one minus sign.
But the Maxwell scalar $\phi_1$ is still remained the same as in
(2.12). Thus, we have shown the changes in the mass of the
Reissner-Nordstrom-de Sitter black hole after every electrical
radiation. Hence, it follows the theorem 1 cited above in the case
of non-rotating variable-charged black hole. This also indicates
the incorporation of Boulware's suggestion [7] that `the
stress-energy tensor may be used to calculate the change in the
metric due to the radiation'. The classical spacetime metrics
(2.16), (2.21) and (2.23) represent the relativistic aspects of
Hawking's radiation in Reissner-Nordstrom-de Sitter black hole

Comparing the metrics (2.16) and (2.23), one observes that the
classical spacetime (2.23) describes a non-rotating spherical
symmetric star with a negative mass ${\cal M}^*$. Such objects
with negative masses are referred to as naked singularities
[1,2,3]. The metric (2.23) can be regarded as a mathematical
representation of the third possibility of Hawking and Israel [3]
in the case of non-rotating singularity. Here it is noted that the
creation of negative mass naked singularity is mainly based on the
continuous electrical radiation of the variable charge $e(r)$ in
the energy-momentum tensor of Einstein-Maxwell equations under the
decomposition (2.1) of Ricci scalar $\Lambda$.

\subsection{\it Variable-charged Kerr-Newman-de Sitter
solution}

Here, we shall incorporate the relativistic aspect of Hawking's
radiation in variable-charged Kerr-Newman-de Sitter black hole
when the electric charge $e$ is taken as a function of $r$ in the
Einstein-Maxwell field equations. The line element with $e(r)$ is
\begin{eqnarray}
ds^2&=&\Big[1-R^{-2}\Big\{2Mr-e^2(r)+{\Lambda^*r^4\over
3}\Big\}\Big]\,du^2+2du\,dr \cr
&&+2aR^{-2}\Big\{2Mr-e^2(r)+{\Lambda^*r^4\over
3}\Big\}\,{\rm sin}^2
\theta\,du\,d\phi-2a\,{\rm sin}^2\theta\,dr\,d\phi \cr
&&-R^2d\theta^2-\Big\{(r^2+a^2)^2
-\Delta^*a^2\,{\rm sin}^2\theta\Big\}\,
R^{-2}{\rm sin}^2\theta\,d\phi^2,
\end{eqnarray}
where
\begin{equation}
R^2=r^2+a^2{\rm cos}^2\theta,\,\;\;
\Delta^*=r^2-2Mr+a^2+e^2(r)-\Lambda^*r^4/3.
\end{equation}
This metric will recover the rotating Kerr-Newman-de Sitter
solution given in appendix (C1), when $e$ becomes constant. The
null tetrad vectors are chosen as
\begin{eqnarray*}
\ell_a=\delta^1_a -a\,{\rm sin}^2\theta\,\delta^4_a,
\end{eqnarray*}
\begin{eqnarray*}
n_a={\Delta^*\over 2\,R^2}\,\delta^1_a+ \delta^2_a
-{\Delta^*\over 2\,R^2}\,a\,{\rm
sin}^2\theta\,\delta^4_a,
\end{eqnarray*}
\begin{equation}
m_a=-{1\over\surd 2R}\,\Big\{-ia\,{\rm
sin}\,\theta\,\delta^1_a+R^2\,\delta^3_a +i(r^2+a^2)\,{\rm
sin}\,\theta\,\delta^4_a\Big\},
\end{equation}
\begin{eqnarray*}
\overline m_a=-{1\over\surd 2\overline R}\,\Big\{ia\,{\rm
sin}\,\theta\,\delta^1_a +R^2\,\delta^3_a-i(r^2+a^2)\,{\rm
sin}\,\theta\,\delta^4_a\Big\}.
\end{eqnarray*}
where $R=r+i\,a\,{\rm cos}\,\theta$. Then we solve the
Einstein-Maxwell field equations for the metric (2.27) and write
only the changed NP quantities
\begin{equation}
\mu=-{1\over{2\overline R\,R^2}}\Big\{r^2-2Mr+a^2+e^2(r)
-{\Lambda^*r^4\over 3}\Big\},
\end{equation}
\begin{eqnarray}
\gamma&=&{1\over{2\overline
R\,R^2}}\,\Big[\Big\{r-M+e(r)\,e'(r)-{\Lambda^*r^4\over 3}
\Big\}\overline R \cr
&&-\Big\{r^2-2Mr+a^2+e^2(r)-{2\Lambda^*r^4\over 3}\Big\}\Big],
\end{eqnarray}
\begin{eqnarray}
\psi_2&=&{1\over{\overline R\,\overline R\,R^2}}\,\Big\{-MR +
e^2(r)-e(r)e'(r)\,{\overline R}+{\Lambda^*r^2\over 3} a^2\,{\rm
cos}^2\theta\Big\} \cr
&&+{1\over{6\,R^2}}\Big\{e'^2(r)+e(r)\,e''(r)\Big\},
\end{eqnarray}
\begin{eqnarray}
\phi_{11}&=&{1\over {2\,R^2\,R^2}}\,\Big\{e^2(r)-2r\,e(r)e'(r)
-{\Lambda^*r^2}\,a^2\,{\rm cos}^2\theta\Big\} \cr && + {1\over
4R^2}\,\Big\{e'^2(r)+e(r)\,e''(r)\Big\},\;\
\end{eqnarray}
\begin{equation}
\Lambda ={\Lambda^*r^2\over 6\,R^2}
-{1\over 12\,R^2}\,\Big\{e'^2(r)+e(r)\,e''(r)\Big\},
\end{equation}
where a prime denotes the derivative with respect to $r$. We have
seen that in each expression of $\phi_{11}$ and $\psi_2$ there is
a cosmological $\Lambda^*$ term coupling with the rotation
parameter $a$. In the case of non-rotating Reissner-Nordstrom-de
Sitter black hole, such $\Lambda^*$ term does not involve in the
expression of $\phi_{11}$ and $\psi_2$ as seen in (2.4) and (2.6).
Hence, without loss of generality, it will be convenient here to
have a decomposition of $\phi_{11}$ into two parts - one for the
cosmological Ricci scalar $\phi_{11}^{(\rm C)}$ and the other for
the electromagnetic  field $\phi_{11}^{(\rm E)}$ as in the case of
$\Lambda$ in (2.1), such that
\begin{equation}
\phi_{11}^{(\rm C)}=-{1\over {2\,R^2\,R^2}}\,
{\Lambda^*r^2}\,a^2\,{\rm cos}^2\theta
\end{equation}
\begin{equation}
\phi_{11}^{(\rm E)}={1\over
{2\,R^2\,R^2}}\,\Big\{e^2(r)-2r\,e(r)e'(r)\Big\}+
{1\over 4R^2}\,\Big\{e'^2(r)+e(r)\,e''(r)\Big\}.
\end{equation}
Similarly, we also have the decomposition of $\Lambda$ as in (2.1)
\begin{equation}
\Lambda^{(\rm C)} ={\Lambda^*r^2\over 6\,R^2}
\end{equation}
\begin{equation}
\Lambda^{(\rm E)} =-{1\over 12\,R^2}\,\Big\{e'^2(r)
+e(r)\,e''(r)\Big\}.
\end{equation}
From (2.8) and (2.35) we have seen the difference between the two
cosmological Ricci scalars $\Lambda^{\rm C}$ of non-rotating and
rotating black holes.

Now, the scalar $\Lambda^{(\rm E)}$ for electromagnetic field must
vanish for this rotating metric. Thus, the vanishing
$\Lambda^{(\rm E)}$ of the equation (2.36) yields that
\begin{equation}
e^2(r) = 2\,rm_1 + C
\end{equation}
where  $m_1$ and $C$ are real constants of integration.
Then, substituting this result in equation (2.34) we obtain the
Ricci scalar for electromagnetic field
\begin{equation}
\phi_{11}^{(\rm E)}={C\over 2\,R^2\,R^2}.
\end{equation}
however, the cosmological Ricci scalar $\phi_{11}^{(\rm C)}$
remains the same form as in (2.33). Accordingly, the Maxwell
scalar takes, after identifying the constant $C\equiv e^2$,
\begin{equation}
\phi_1={e\over\surd 2\,\overline R\,\overline R}.
\end{equation}
Hence, the changed NP quantities
\begin{equation}
\mu=-{1\over{2\overline R\,R^2}}\Big\{r^2-2r(M-m_1)+a^2
+e^2-{\Lambda^*r^4\over 3}\Big\},
\end{equation}
\begin{eqnarray*}
\gamma={1\over{2\overline
R\,R^2}}\,\Big[\Big\{r-(M-m_1)\Big\}\overline R
-\Big\{r^2-2r(M-m_1)
+a^2+e^2-{\Lambda^*r^4\over 3}\Big\}\Big],
\end{eqnarray*}
\begin{equation}
\psi_2={1\over{\overline R\,\overline R\,R^2}},
\Big\{-(M-m_1)R + e^2+{\Lambda^*r^2\over3}\,a^2\,{\rm
cos}^2\theta\Big\},
\end{equation}
\begin{equation}
\phi_{11}={1\over {2\,R^2\,R^2}}\,\Big\{e^2(r)
-{\Lambda^*r^2}\,a^2\,{\rm cos}^2\theta\Big\},
\end{equation}
\begin{equation}
\Lambda=\Lambda^{(\rm C)} ={\Lambda^*r^2\over 6\,R^2}.
\end{equation}
We have seen the changes in $\mu$, $\gamma$ and $\psi_2$, but no
changes in $\phi_{11}$ and $\Lambda$. Thus, the rotating solution
(2.25) with a new constant $m_1$ after the first radiation becomes
\begin{eqnarray}
ds^2&=&\Big[1-R^{-2}\Big\{2r(M-m_1)-e^2+{\Lambda^*r^4\over
3}\Big\}\Big]\,du^2
+2du\,dr \cr
&&+2aR^{-2}\Big\{2r(M-m_1)-e^2+{\Lambda^*r^4\over
3}\Big\}\,{\rm sin}^2
\theta\,du\,d\phi-2a\,{\rm sin}^2\theta\,dr\,d\phi \cr
&&-R^2d\theta^2-\Big\{(r^2+a^2)^2
-\Delta^*a^2\,{\rm sin}^2\theta\Big\}\,
R^{-2}{\rm sin}^2\theta\,d\phi^2,
\end{eqnarray}
where
\begin{equation}
\Delta^*=r^2-2r(M-m_1)+a^2+e^2-\Lambda^*r^4/3.
\end{equation}
This suggests that the first electrical radiation of rotating
black hole leads to a reduction of the gravitational mass $M$ by a
quantity $m_1$ with the same Maxwell scalar $\phi_1$ and the
constant $\Lambda^{(\rm C)}$. If we consider another radiation by
taking $e$ in (2.44) to be a function of $r$ with the mass $M-m_1$
and the decomposition (2.1), then the Einstein-Maxwell field
equations yield to reduce this mass by another constant quantity
$m_2$ (say); i.e., after the second radiation, the mass will
become $M-(m_1+m_2)$. Here again, the  Maxwell scalar $\phi_1$ and
the constant $\Lambda^{(\rm C)}$ remain the same form after the
second radiation also. Thus, if we consider $n$ radiations
everytime taking the charge $e$ to be function of $r$ with the
decomposition of $\Lambda$, the Maxwell scalar $\phi_1$ will be
the same, but the metric takes the form:
\begin{eqnarray}
ds^2&=&\Big[1-R^{-2}\Big\{2r{\cal M}-e^2+{\Lambda^*r^4\over
3}\Big\}\Big]\,du^2+2du\,dr \cr
&&+2aR^{-2}\Big\{2r{\cal M}-e^2+{\Lambda^*r^4\over
3}\Big\}\,{\rm sin}^2
\theta\,du\,d\phi-2a\,{\rm sin}^2\theta\,dr\,d\phi \cr
&&-R^2d\theta^2-\Big\{(r^2+a^2)^2
-\Delta^*a^2\,{\rm sin}^2\theta\Big\}\,R^{-2}{\rm
sin}^2\theta\,d\phi^2,
\end{eqnarray}
where the total mass of the black hole, after the $n$ radiations
will be of the form
\begin{equation}
{\cal M}=M-(m_1 + m_2 + m_3 + m_4 + . . .+ m_n).
\end{equation}
Taking Hawking's radiation of black holes, we can expect that the
total mass of  black hole will be radiated away just leaving
${\cal M}$ equivalent to Planck mass of about $10^{-5}$ g, that
is, $M$ may not be exactly equal to $m_1 + m_2 + m_3 + m_4 + . .
.+ m_n$, but has a difference of about Planck-size mass, as in the
case of non-rotating black hole. Otherwise, the total mass of
black hole will be evaporated completely after continuous
radiation when ${\cal M} = 0$, that is, $M = m_1 + m_2 + m_3 + m_4
+ . . .+ m_n$. Here the rotating variable-charged black hole might
completely radiate away its mass just leaving the electrical
charge $e$ and the cosmological constant $\Lambda^*$. We find this
situation in the form of classical space-time metric as
\begin{eqnarray}
ds^2&=&\Big\{1+\Big(e^2-{\Lambda^*r^4\over
3}\Big)\,R^{-2}\Big\}\,du^2+2du\,dr \cr
&&-2a\,R^{-2}\,\Big(e^2-{\Lambda^*r^4\over
3}\Big)\,{\rm sin}^2
\theta\,du\,d\phi-2a\,{\rm sin}^2\theta\,dr\,d\phi \cr
&&-R^2d\theta^2-\Big\{(r^2+a^2)^2
-\Delta^*a^2\,{\rm sin}^2\theta\Big\}\,R^{-2}{\rm
sin}^2\theta\,d\phi^2,
\end{eqnarray}
with the charge $e$ and the cosmological constant $\Lambda^*$, but
no mass, where $\Delta^*=r^2+a^2+e^2-\Lambda^*r^4/3$. The metric
(2.48) will describe a rotating `instantaneous' naked singularity
with zero mass in de Sitter space. At this stage, the Weyl scalar
$\psi_2$ takes the form
\begin{equation}
\psi_2={1\over{\overline R\,\overline R\,R^2}}\Big\{
e^2+{\Lambda^*r^2\over3}\,a^2\,{\rm cos}^2\theta\Big\}\;\
\end{equation}
showing the gravity on the surface of the remaining solution
depending on the electric charge $e$ and the cosmological constant
$\Lambda^*$ coupling with the rotational parameter $a$; however,
the Maxwell scalar $\phi_1$ and the Ricci scalar $\Lambda^{(\rm
C)}$ remain the same as in (2.39) and (2.35) respectively. This
completes the other part of the theorem 2 cited above.

It means that there may be rotating black holes in the universe
whose masses are completely radiated; their gravity depends on the
electric charge of the body and the cosmological constant
$\Lambda^*$, and their metrics appear similar to that in (2.48).
Here the idea of this complete evaporation of masses of radiating
black holes embedded in the de Sitter space is in agreement with
that of Hawking's evaporation of black holes. It is worth studying
the nature of such rotating black holes (2.48) or in the
case of non-rotating (2.21). This might give a different physical
nature, which one has not been studied in common theory of black
holes embedded into the de Sitter space.

   Here, we again consider the  charge $e$ to be function
of radial coordinate $r$ for next radiation in (2.48), so that we
get, from the Einstein's field equations, the scalar
$\Lambda^{(\rm E)}$ as given in equation (2.36) and the same
scalar $\Lambda^{(\rm C)}$ as in (2.35). Then the vanishing of
this $\Lambda^{(\rm E)}$ for electromagnetic field will lead to
create a new mass (say $m^*_1$) in the remaining space-time
geometry (2.48). For the second radiation, we again consider the
charge $e$ to be function of $r$ in the field equations with the
mass $m^*_1$. Then the vanishing of $\Lambda^{(\rm E)}$ will lead
to increase the new mass by another quantity $m^*_2$ (say) i.e.,
after the second radiation of (2.48) the new mass will be $m^*_1
+m^*_2$. If this radiation process continues further for a long
time, the new mass will increase gradually as
\begin{equation}
{\cal M}^*=m^*_1 + m^*_2 + m^*_3 + m^*_4 + . . .  .\,\,
\end{equation}
Then the spacetime metric will take the form
\begin{eqnarray}
ds^2&=&\Big[1+R^{-2}\Big\{2r{\cal M}^*+e^2-{\Lambda^*r^4\over
3}\Big\}\Big]\,du^2+2du\,dr \cr
&&-2aR^{-2}\Big\{2r{\cal M}^*+e^2-{\Lambda^*r^4\over
3}\Big\}\,{\rm sin}^2
\theta\,du\,d\phi-2a\,{\rm sin}^2\theta\,dr\,d\phi \cr
&&-R^2d\theta^2-\Big\{(r^2+a^2)^2
-\Delta^*a^2\,{\rm sin}^2\theta\Big\}\,
R^{-2}{\rm sin}^2\theta\,d\phi^2,
\end{eqnarray}
where $\Delta^*=r^2+2r{\cal M}^*+a^2+e^2-{\Lambda^*r^4/3}$. The
changed NP quantities $\psi_2$ and $\mu$ are
as follows
\begin{equation}
\psi_2={1\over{\overline R\,\overline R\,R^2}}\,\Big\{{\cal
M}^*R+e^2+{\Lambda^*r^2\over3}\,a^2\,{\rm cos}^2\theta\}.\;\
\end{equation}
\begin{equation}
\mu=-{1\over{2\overline R\,R^2}}\Big\{r^2+2r{\cal
M}^*+a^2+e^2-{\Lambda^*r^4\over 3}\Big\},
\end{equation}
with $\phi_1$ given in (2.39) and $\Lambda^{(\rm C)}$ in (2.35).
Comparing the metrics (2.46) and (2.51), we find that the
classical spacetime (2.51) describes a rotating negative mass
naked singularity embedded into the de Sitter cosmological space.
Thus, from the above it follows the proof of the rotating part of
theorem 3. We have also shown the possible changes in the mass of
the rotating charged de Sitter black hole without affecting the
Maxwell scalar $\phi_1$ as well as the cosmological constant
$\Lambda^*$, and accordingly, metrics are cited for future use.
Thus, this completes the proof of other part of the theorem 1 for
the rotating charged de Sitter black hole. Also since there is no
effect on the cosmological constant $\Lambda^*$ during Hawking's
evaporation process, it will always remain unaffected. That is,
unless some external forces apply to remove the cosmological
constant $\Lambda^*$ from the spacetime geometries, it will
continue to exist along with the electrically radiating objects,
rotating or non-rotating forever. This leads to the proof of the
theorem 4 cited above for both the rotating as well as
non-rotating black holes.

\section{Conclusion}
\setcounter{equation}{0}
\renewcommand{\theequation}{3.\arabic{equation}}

In the above section, we have shown that the Hawking's radiation
can be expressed in classical spacetime metrics, by considering
the charge $e$ to be the function of the radial coordinate $r$ of
Reissner-Nordstrom-de Sitter as well as Kerr-Newman-de Sitter
black holes. That is, every electrical radiation produces a change
in the mass of the charged objects. These changes in the mass of
black holes embedded in de Sitter space, after every electrical
radiation, describe the relativistic aspect of Hawking's
evaporation of masses of black holes in the classical spacetime
metrics. This follows from the statement of theorem 1. It appears
that the black hole evaporation in de Sitter space is mainly based
on the decomposition of the Ricci scalar $\Lambda$ into two parts
- one for the cosmological object and other for the
electromagnetic field. That is, the Ricci scalar for the
cosmological de Sitter space is generally non-vanishing whereas
that for the electromagnetic field is always vanished. Thus the
decomposition of Ricci scalar $\Lambda$ for the black holes
embedded in de Sitter space is, without loss of generality be
possible, because the cosmological objects and the electromagnetic
fields are two different matter fields possessing different
physical properties. So the vanishing Ricci scalar for
electromagnetic field with the charge $e(r)$ makes the change in
the mass of black holes embedded in the de Sitter space. Thus, we
find that the black hole radiation process in de Sitter space is
based on the electrical radiation of the variable-charge $e(r)$ in
the energy-momentum tensor describing the change in mass in
classical spacetime metrics which is in agreement with Boulware's
suggestion [7] mentioned above. The Hawking's evaporation of
masses and the creation of negative mass naked singularities are
also due to the continuous electrical radiation. The formation of
naked singularity of negative mass is also Hawking's suggestion
[2] mentioned in the introduction above. This suggests that, if
one accepts the continuous electrical radiation to lead the
complete evaporation of the original mass of black holes, then the
same radiation will also lead to the creation of new mass to form
negative mass naked singularities embedded into the de Sitter
space. This clearly suggests that an electrically radiating black
hole, rotating or non-rotating will not disappear completely,
which is against the suggestion made in [1,2,3,5]. The
disappearance of such a black hole during the radiation process
will be for an instant, thereby the formation of `instantaneous'
de Sitter naked singularities. This follows the statement of the
theorem 2 above for both rotating and  non-rotating black holes.

Now the spacetime metric (2.51) representing the rotating
negative mass naked singularity in de Sitter space can be written
in the Boyer-Lindquist coordinates $(t,r,\theta,\phi)$ for future
use as
\begin{eqnarray}
ds^2&=&\Big[1+R^{-2}\Big\{2r\,{\cal M}^*+e^2-{\Lambda^*r^4\over
3}\Big\}\Big]\,dt^2-{R^2\over\Delta^*}dr^2
-R^2d\theta^2 \cr
&&-\{(r^2+a^2)^2
-\Delta^*a^2\,{\rm sin}^2\theta\}\,R^{-2}{\rm
sin}^2\theta\,d\phi^2 \cr
&&-2a\Big(2r\,{\cal M}^*+e^2-{\Lambda^*r^4\over
3}\Big)\,R^{-2}\,{\rm sin}^2\theta\,dt\,d\phi.
\end{eqnarray}
where $\Delta^* = r^2+2r\,{\cal M}^*+a^2+e^2-{\Lambda^*r^4/3}$.

The metric (2.51) describing rotating negative mass naked
singularity can be written in two Kerr-Schild forms on different
backgrounds as follows:
\begin{equation}
g_{ab}^{\rm NMdS}=g_{ab}^{\rm dS} +2Q(r,\theta)\ell_a\ell_b
\end{equation}
where $Q(r,\theta) =(r{\cal M}^*+e^2/2)R^{-2}$,
and
\begin{eqnarray}
g_{ab}^{\rm NMdS}=g_{ab}^{\rm
NM}+2Q(r,\theta)\,\ell_a\,\ell_b,
\end{eqnarray}
with $Q(r,\theta)=-(\Lambda^*r^4/6)R^{-2}$. $g_{ab}^{\rm NM}$ is
the metric tensor for rotating negative mass naked singularity
whose the line element is given below
\begin{eqnarray}
ds^2&=&\Big[1+R^{-2}\Big(2r{\cal M}^*+e^2
\Big)\Big]\,du^2+2du\,dr \cr
&&-2aR^{-2}\Big(2r{\cal M}^*+e^2\Big)\,{\rm sin}^2
\theta\,du\,d\phi-2a\,{\rm sin}^2\theta\,dr\,d\phi \cr
&&-R^2d\theta^2-\Big\{(r^2+a^2)^2
-\Delta^*a^2\,{\rm sin}^2\theta\Big\}\,
R^{-2}{\rm sin}^2\theta\,d\phi^2,
\end{eqnarray}
where $\Delta^*=r^2+2r{\cal M}^*+a^2+e^2$. The Kerr-Schild (3.2)
provides the classical spacetime metric representing negative mass
naked singularity embedded into the rotating de Sitter universe
and the Kerr-Schild form (3.3) shows that the rotating de Sitter
space is embedded into the negative mass naked singularity
background. The null vector $\ell_a$, appearing in (3.2) and (3.3)
is one of the repeated principal null directions of $g_{ab}^{\rm
NMdS}$, $g_{ab}^{\rm NM}$ and $g_{ab}^{\rm dS}$ which are all
Petrov type $D$ spacetime metrics. Also $\ell_a$ is geodesic,
shear free, expanding as well as non-zero twist null vector.

As the negative mass naked singularity metric $g_{ab}^{\rm NMdS}$
(2.51) describes Petrov $D$ spacetime, it is worth mentioning that
Chandrasekhar [15] has established a relation of spin coefficients
$\rho$, $\mu$, $\tau$, $\pi$ in the case of an affinely
parameterized geodesic vector, generating an integral which is
constant along the geodesic in a vacuum Petrov type $D$
space-time
\begin{equation}
{\rho\over \overline\rho}={\mu\over\overline\mu}={\tau\over\overline\pi}
={\pi\over\overline\tau}.
\end{equation}
This relation has been derived on the  basis of the vacuum Petrov
type $D$ space-time with $\psi_2\neq 0$, $\psi_0=\psi_1
=\psi_3=\psi_4=0$ and $\phi_{01}=\phi_{02}=\phi_{10}
=\phi_{20}=\phi_{12}=\phi_{21}=\phi_{00}=\phi_{22}=\phi_{11}=\Lambda=0$.
In [16] for non-vacuum Petrov type $D$ spacetimes, the
Killing-Yano (KY) scalar $\chi_1={1\over 2}
f_{ab}(\ell^a\,n^b+\overline{m}^a\,m^b)$ has been introduced in
(3.5) as
\begin{equation}
{\rho\over \overline\rho}={\mu\over\overline\mu}={\tau\over\overline\pi}
={\pi\over\overline\tau}=-{\overline\chi_1\over\chi_1},
\end{equation}
where $f_{ab}$ is KY tensor satisfying the KY equations
\begin{equation}
f_{ab;c}+f_{ac;b}=0.
\end{equation}
The relation (3.6) can also be found in equation (31.63) of
Kramer, et al. [17], if one puts $\chi_1=A+iB$ with a little
calculation. The importance of KY tensor in General Relativity
seems to lie on Carter's remarkable result [18] that the
separation constant of Hamilton-Jacobi equation (for charged
orbits) in the Kerr space-time gives a fourth constant. In fact,
this constant arises from the scalar field $K_{ab}v^a\,v^b$ which
has vanishing divergent along a unit vector $v^a$ tangent to an
orbit of charged particle. Here $K_{ab} =f_{ma}f^m_b$.

It is emphasized that the KY equations in NP formalism given in
[16] for non-vacuum Petrov type $D$ can be solved for the
negative mass naked singularity embedded into the  de Sitter
spacetime metric (2.51) to obtain the KY scalar $\chi_1$ as
\begin{equation}
\chi_1=iC(r-i\,a\,{\rm cos}\,\theta),
\end{equation}
where $C$ is a real constant. Ultimately, the relation (3.6) takes
the form
\begin{equation}
{\rho\over \overline\rho}={\mu\over\overline\mu}={\tau\over\overline\pi}
={\pi\over\overline\tau}=-{\overline\chi_1\over\chi_1}
={R\over\overline R},
\end{equation}
where $R=r+i\,a\,{\rm cos}\,\theta$ and the spin coefficients
$\rho$, $\mu$, $\tau$, $\pi$ are unchanged quantities for the
negative mass naked singularity except $\mu$. So they are
presented in the appendix C and $\mu$ is given in (2.53) above.
Thus, it concludes that the negative mass naked singularity embedded into
the de Sitter cosmological space satisfies the  extended version
of Chandrasekhar's relation (3.6).

We have seen from above that the changes in the masses of
Reissner-Nordstrom-de Sitter as well as Kerr-Newman-de Sitter
black holes take place due to the vanishing of Ricci scalar
$\Lambda^{(\rm E)}$ and both the Maxwell scalar $\phi_1$ and
cosmological Ricci scalar $\Lambda^{(\rm C)}$ remained unchanged
in the field equations for a variable charge. So, if the Maxwell
scalar $\phi_1$ is absent from the spacetime geometry, there will
be no electrical radiation and, consequently, no change in the
mass of the black hole will take place in the black hole.
Therefore, one cannot expect, in principle to observe such
relativistic change in the masses  in the case of uncharged
Schwarzschild-de Sitter as well as Kerr-de Sitter black holes.
Taking Hawking radiation into account, it emphasizes that the time
taken of one radiation to another will be very short that one may
not, practically realize after losing $m_1$ from the mass how
quickly $m_2$ is being reduced  and so on, as seen above in
Section 2. It is found that the metric (2.21) or (2.48) without
mass will occur {\sl only} for a very short period, as the
electrical radiation continues further. Thus, we have incorporated
Hawking radiation in relativistic viewpoint in classical curved
spacetime geometries which will describe the possible life style
of electrically radiating black holes embedded into the de Sitter
universe during their radiation process. One might expect that the
metrics with mass ${\cal M}^*$ in equations (2.23) and (2.51)
might have a different physical feature to the
Reissner-Nordstrom-de Sitter as well as Kerr-Newman-de Sitter
solutions.

\section*{Acknowledgement}

The author acknowledges his appreciation for hospitality received
from Inter-University Centre for Astronomy and Astrophysics
(IUCAA), Pune during the preparation of the paper.

\section*{Appendix}

Here we shall present the rotating de Sitter,
Reissner-Nordstrom-de Sitter as well as Kerr-Newman-de Sitter
metrics in NP formalism for a quick reference. From the rotating
Kerr-Newman-de Sitter solution ($M\neq0,a\neq0,e\neq0$), the
rotating de Sitter ($M=0,a\neq0,e=0$) and Reissner-Nordstrom-de
Sitter ($M\neq0,a=o,e\neq0$) metrics can be recovered. It is also
worth mentioning that from the rotating Kerr-Newman-de Sitter
metric, we can recover a rotating charged de Sitter cosmological
universe ($M=0,a\neq 0,e\neq0$) which has been referred in section 2. \\\\\
A. {\it Rotating de Sitter solution}:  \\\\\
\setcounter{equation}{0}
\renewcommand{\theequation}{A\arabic{equation}}
The rotating de Sitter solution is derived by using Wang-Wu
function [20] in general rotating solutions in [12]. The line
element is given as
\begin{eqnarray}
ds^2&=&\Big\{1-{\Lambda^*\,r^4\over3\,R^2}
\Big\}\,du^2
+2du\,dr \cr
&&+2a{\Lambda^*\,r^4\over3}\,R^{-2}\,{\rm
sin}^2\theta\,du\,d\phi-2a\,{\rm sin}^2\theta\,dr\,d\phi \cr
&&-R^2d\theta^2-\Big\{(r^2+a^2)^2
-\Delta^*a^2\,{\rm sin}^2\theta\Big\}\,R^{-2}{\rm
sin}^2\theta\,d\phi^2,
\end{eqnarray}
where $R^2=r^2+a^2{\rm cos}^2\theta$,
$\Delta^*=r^2-{\Lambda^*\,r^4}/3+a^2$.
Here $\Lambda^*$ denotes the cosmological constant of the de
Sitter space.
The null tetrad vectors are
\begin{eqnarray*}
\ell_a=\delta^1_a -a\,{\rm sin}^2\theta\,\delta^4_a,
\end{eqnarray*}
\begin{eqnarray*}
n_a={\Delta^*\over 2\,R^2}\,\delta^1_a+ \delta^2_a
-{\Delta^*\over 2\,R^2}\,a\,{\rm
sin}^2\theta\,\delta^4_a,
\end{eqnarray*}
\begin{equation}
m_a=-{1\over\surd 2R}\,\Big\{-ia\,{\rm
sin}\,\theta\,\delta^1_a+R^2\,\delta^3_a +i(r^2+a^2)\,{\rm
sin}\,\theta\,\delta^4_a\Big\},
\end{equation}
\begin{eqnarray*}
\overline m_a=-{1\over\surd 2\overline R}\,\Big\{ia\,{\rm
sin}\,\theta\,\delta^1_a +R^2\,\delta^3_a-i(r^2+a^2)\,{\rm
sin}\,\theta\,\delta^4_a\Big\}.
\end{eqnarray*}
where $R=r+i\,a\,{\rm cos}\theta$. Then the NP quantities are
\begin{eqnarray*}
\kappa=\sigma=\nu=\lambda=\epsilon=0,
\end{eqnarray*}
\begin{eqnarray}
\rho&=&-{1\over\overline R},\;\;\;\;
\mu=-{\Delta^*\over{2\overline R\,R^2}},\:\:\:
\alpha={(2ai-R\,{\rm cos}\,\theta)\over{2\surd 2\overline
R\,\overline R\,{\rm sin}\,\theta}},                  \cr
\beta&=&{{\rm cot}\,\theta\over2\surd 2R},\:\:\: \pi={i\,a\,{\rm
sin}\,\theta\over{\surd 2\overline R\,\overline R}},\;\;\;
\tau=-{i\,a\,{\rm sin}\,\theta\over{\surd 2R^2}},
\end{eqnarray}
\begin{eqnarray*}
\gamma=-{1\over{2\overline R\,R^2}}\,\left\{(1-{1\over
3}\Lambda^*r^2)r\,\overline R+\Delta^*\right\},
\end{eqnarray*}
\begin{equation}
\phi_{11}= -{1\over {2\,R^2\,R^2}}\Lambda^*r^2a^2\,{\rm
cos}^2\theta,\;\;\;
\psi_2={1\over3{\overline R\,\overline
R\,R^2}}\Lambda^*r^2a^2{\rm
cos}^2\theta,
\end{equation}
\begin{equation}
\Lambda={\Lambda^*r^2\over{6R^2}}.
\end{equation}
From these NP quantities we certainly observe that the
rotating de Sitter metric is a Petrov type $D$
gravitational field, whose one of the repeated principal null
vectors, $\ell_a$ is geodesic, shear free, expanding as well as
rotating. The rotating cosmological space possesses an
energy-momentum tensor
\begin{eqnarray}
T_{ab} =2\,\rho^*\,\ell_{(a}\,n_{b)}
+2\,p\,m_{(a}\overline m_{b)},
\end{eqnarray}
where $K\,\rho^*=2\,\phi_{11} + 6\,\Lambda$ and
$K\,p=2\,\phi_{11} - 6\,\Lambda$. are related to the density and
the pressure  of the cosmological matter
which is, however not a perfect fluid. \\\\\
B. {\it Reissner-Nordstrom-de Sitter solution}: \\\\\
\setcounter{equation}{0}
\renewcommand{\theequation}{B\arabic{equation}}
This is a spherically symmetric non-rotating solution
\begin{equation}
ds^2=(1-{2M\over r}+{e^2\over r^2}-{\Lambda^*r^2\over
3})\,du^2+2du\,dr-r^2d\theta^2
+r^2\,{\rm sin}^2\theta\,d\phi^2,
\end{equation}
where $M$ and $e$ are the mass and charge respectively. For this
metric one chooses the null tetrad vectors
the covariant complex null tetrad vectors take the forms
\begin{equation}
\ell_a=\delta^1_a,\;\;\;
n_a={1\over 2}\Big\{1-{2m\over
r}+{e^2\over r^2}-{\Lambda^*r^2\over
3}\Big\}\delta^1_a+\delta^2_a,
\end{equation}
\begin{eqnarray*}
m_a=-{r\over\surd 2}\,\{\delta^3_a+i\,{\rm
sin}\,\theta\,\delta^4_a\}.
\end{eqnarray*}
The NP quantities are
\begin{eqnarray*}
\kappa=\sigma=\nu=\lambda=\pi=\tau=\epsilon=0,\;\;\ \rho=-{1\over
r},\;\ \beta=-\alpha={1\over {2\surd 2r}}\,{\rm cot}\,\theta,
\end{eqnarray*}
\begin{equation}
\mu=-{1\over 2\,r}\Big(1-{2M\over r}+{e^2\over
r^2}-{\Lambda^*r^2\over
3}\Big),\:\:
\gamma={1\over 2\,r^3}\,(M\,r-e^2),
\end{equation}
\begin{equation}
\psi_2=-(M\,r-e^2)\,r^{-4},\;\;
\phi_{11}={1\over2}\,e^2\,r^{-4},\;\;\;
\Lambda={\Lambda^*\over6}
\end{equation}
We have seen that there is no $\Lambda^*$ terms in $\phi_{11}$ and
$\psi_2$ for this non-rotating metric. The energy momentum tensor
for the Reissner-Nordstrom-de Sitter metric
takes the form
\begin{equation}
T_{ab}=T_{ab}^{(\rm E)}+\Lambda^*g_{ab},
\end{equation}
where the energy-momentum tensor for the non-null electromagnetic
field is given as
\begin{equation}
T_{ab}^{(\rm E)}={4\,e^2\over K\,r^4}\{\ell_{(a}\,n_{b)}
+m_{(a}\overline m_{b)}\}
\end{equation}
The EMT (B5) can be treated as Guth's modification of energy
momentum tensor [19] for electromagnetic field in early
inflationary universe.\\\\\
C. {\it Kerr-Newman-de Sitter solution}: \\\\\
\setcounter{equation}{0}
\renewcommand{\theequation}{C\arabic{equation}}
The line element of the Kerr-Newman-de Sitter solution is given in
[12] as
\begin{eqnarray}
ds^2&=&\Big\{1-R^{-2}\Big(2Mr-e^2
+{\Lambda^*\,r^4\over3}\Big)
\Big\}\,du^2
+2du\,dr \cr
&&+2aR^{-2}\Big(2Mr-e^2+{\Lambda^*\,r^4\over 3}\Big)\,{\rm
sin}^2\theta\,du\,d\phi-2a\,{\rm sin}^2\theta\,dr\,d\phi \cr
&&-R^2d\theta^2-\Big\{(r^2+a^2)^2
-\Delta^*a^2\,{\rm sin}^2\theta\Big\}\,R^{-2}{\rm
sin}^2\theta\,d\phi^2,
\end{eqnarray}
where $R^2=r^2+a^2{\rm cos}^2\theta$,
$\Delta^*=r^2-2Mr-{\Lambda^*\,r^4}/3+a^2+e^2$.
The null tetrad vectors are
\begin{eqnarray*}
\ell_a=\delta^1_a -a\,{\rm sin}^2\theta\,\delta^4_a,
\end{eqnarray*}
\begin{equation}
n_a={\Delta^*\over 2\,R^2}\,\delta^1_a+ \delta^2_a
-{\Delta^*\over 2\,R^2}\,a\,{\rm
sin}^2\theta\,\delta^4_a,
\end{equation}
\begin{eqnarray*}
m_a=-{1\over\surd 2R}\,\Big\{-ia\,{\rm
sin}\,\theta\,\delta^1_a+R^2\,\delta^3_a +i(r^2+a^2)\,{\rm
sin}\,\theta\,\delta^4_a\Big\},
\end{eqnarray*}
where $R=r+i\,a\,{\rm cos}\theta$. Then the NP quantities are
\begin{eqnarray*}
\kappa=\sigma=\nu=\lambda=\epsilon=0,
\end{eqnarray*}
\begin{eqnarray}
\rho&=&-{1\over\overline R},\;\;\;\;
\mu=-{\Delta^*\over{2\overline R\,R^2}},\;\;\;\;
\alpha={(2ai-R\,{\rm cos}\,\theta)\over{2\surd 2\overline
R\,\overline R\,{\rm sin}\theta}},                 \cr
\beta&=&{{\rm cot}\,\theta\over2\surd 2R},\:\:\: \pi={i\,a\,{\rm
sin}\,\theta\over{\surd 2\overline R\,\overline R}},\;\;\;
\tau=-{i\,a\,{\rm sin}\,\theta\over{\surd 2R^2}},
\end{eqnarray}
\begin{eqnarray*}
\gamma={1\over{2\overline R\,R^2}}\,\left[(r-M-{2\over
3}\Lambda^*r^3)\overline R-\Delta^*\right],
\end{eqnarray*}
\begin{equation}
\psi_2={1\over{\overline R\overline R\,R^2}}\Big\{e^2-M\,R
+{\Lambda^*r^2\over3}a^2{\rm cos}^2\theta\Big\},
\end{equation}
\begin{equation}
\phi_{11}= {1\over {2\,R^2\,R^2}}\Big(e^2-\Lambda^*r^2a^2\,{\rm
cos}^2\theta\Big),
\end{equation}
\begin{equation}
\Lambda={\Lambda^*r^2\over{6R^2}}.
\end{equation}
The energy-momentum tensor for the Kerr-Newman-de Sitter solution
can be written in the form
\begin{eqnarray}
T_{ab} &=& T^{(\rm E)}_{ab} + T^{(\rm C)}_{ab},
\end{eqnarray}
\begin{eqnarray*}
&=&2\,\rho^{*(\rm E)}\{\ell_{(a}\,n_{b)} +m_{(a}\overline m_{b)}\}
+ 2\{\rho^{*(\rm C)}\,\ell_{(a}\,n_{b)} + p^{(\rm
C)}\,m_{(a}\overline m_{b)}\},
\end{eqnarray*}
where the energy densities and pressures for the electromagnetic
field as well as the cosmological matter field are given below:
\begin{eqnarray*}
\rho^{*(\rm E)}&=&p^{(\rm E)}= {e^2\over
{K\,R^2\,R^2}}, \\\\
\rho^{*(\rm C)}&=& {\Lambda^*r^4\over
{K\,R^2\,R^2}}, \:\:\:\:
p^{(\rm C)}= {-\Lambda^*r^2\over{K\,R^2\,R^2}}
\Big(r^2+2a^2\,{\rm cos}^2\theta\Big).
\end{eqnarray*}
From this it is observed that the energy-momentum tensor (C7) will
represent Guth's modification form of $T_{ab}$ [19] for the
rotating metric (C1) with $a\neq0$, as it will take the form (B5)
for non-rotating metric with $a=0$.

One can, without loss of generality, have a decomposition
of the Ricci scalar $\Lambda=g_{ab}R^{ab}$ (C6) into
two parts -- one for the non-zero cosmological scalar
$\Lambda^{(\rm C)}$ and other for the zero Ricci scalar
$\Lambda^{(\rm E)}$ of electromagnetic field as
\begin{equation}
\Lambda=\Lambda^{(\rm C)}+\Lambda^{(\rm E)},
\end{equation}
such that
\begin{equation}
\Lambda^{(\rm C)}={\Lambda^*r^2\over{6R^2}} \:\:\: {\rm and}
\:\:\: \Lambda^{(\rm E)}=0.
\end{equation}
This decomposition of $\Lambda$ has been used in the section 2
above. The derivation of these rotating de Sitter solution and
Kerr-Newman-de Sitter black hole can be seen in [12]. The metric
(C1) can be written in Kerr-Schild form  on the de Sitter
cosmological background space as
\begin{equation}
g_{ab}^{\rm KNdS}=g_{ab}^{\rm dS} +2Q(r,\theta)\ell_a\ell_b
\end{equation}
where $Q(r,\theta) =-(rm-e^2/2)R^{-2}$, and the vector $\ell_a$ is
a geodesic, shear free, expanding as well as rotating null vector
of both $g_{ab}^{\rm KNdS}$ as well as $g_{ab}^{\rm dS}$ and
$g_{ab}^{\rm KN}$ is the Kerr-Newman metric with $m=e$ = constant.
This vector $\ell_a$ is one of the double repeated principal null
vectors of the Weyl tensor of $g_{ab}^{\rm KNdS}$ and $g_{ab}^{\rm
dS}$. The Kerr-Schild form (C10) can also be expressed on the
Kerr-Newman background space as
\begin{eqnarray}
g_{ab}^{\rm KNdS}=g_{ab}^{\rm KN}+2Q(r,\theta)\,\ell_a\,\ell_b,
\end{eqnarray}
where $Q(r,\theta)=-(\Lambda^*r^4/6)\,R^{-2}$, and $\Lambda^*$ is
the cosmological constant and $g_{ab}^{\rm KN}$ is the Kerr-Newman
metric.

\end{document}